\documentclass[aps,pra,amsmath,amssymb,floatfix,twocolumn,amsmath,superscriptaddress,twocolumn,nofootinbib,tighten,letterpaper]{revtex4-2}
\usepackage{multirow}
\usepackage{subfigure}
\usepackage{color}
\usepackage{mathrsfs}
\usepackage{hyperref}
\usepackage[normalem]{ulem}
\usepackage{bm}

\usepackage{amssymb}   
\usepackage{amsmath}

\usepackage{amsfonts, relsize, color}
\usepackage{graphics}
\usepackage{graphicx}
\usepackage{subfigure}
\usepackage{hyperref}
\usepackage{color}
\usepackage{comment}

\newcommand{\bra}[1]{\left< #1 \right|}
\newcommand{\ket}[1]{\left| #1 \right>}

\begin{document}

\title{Dynamic melting and condensation of topological dislocation modes}

\author{Sanjib Kumar Das}
\affiliation{Department of Physics, Lehigh University, Bethlehem, Pennsylvania, 18015, USA}

\author{Bitan Roy}
\affiliation{Department of Physics, Lehigh University, Bethlehem, Pennsylvania, 18015, USA}

\date{\today}

\begin{abstract}
Bulk dislocation lattice defects are instrumental in identifying translationally active topological insulators (TATIs), featuring band inversion at a finite momentum (${\bf K}_{\rm inv}$). As such, TATIs host robust gapless modes around the dislocation core, when the associated Burgers vector ${\bf b}$ satisfies ${\bf K}_{\rm inv} \cdot {\bf b}=\pi$ (modulo $2 \pi$). From the time evolution of appropriate density matrices, we show that when a TATI via a real time ramp enters into a trivial or translationally inert topological insulating phase, devoid of gapless dislocation modes, the signatures of the preramp defect modes survive for a long time. More intriguingly, as the system ramps into a TATI phase from any translationally inert insulator, signature of the dislocation mode dynamically builds up near its core, which is prominent for slow ramps. We exemplify these generic outcomes for two-dimensional time-reversal symmetry breaking insulators. Proposed dynamic responses at the dislocation core can be experimentally observed in quantum crystals, optical lattices and metamaterials with time a tunable band gap.        
\end{abstract}

\maketitle

\section{Introduction and Background}

Interfaces of quantum materials serve as a powerful tool to identify topological crystals in nature. They feature robust gapless modes at the edges and surfaces, for example, encoding the topological invariant of the bulk electronic wavefunctions, manifesting a bulk boundary correspondence~\cite{Hasan-Kane-RMP, Qi-Zhang-RMP}. Here we solely focus on topological insulators (TIs). The hallmark band inversion in TIs, however, can take place at the center ($\Gamma$ point) or at finite time reversal invariant momentum points of the Brillouin zone (BZ)~\cite{liangfu:crystalline, juricic-natphys:crystalline, shiozakisato, slagerkane, fang:crystalline, bernevig:crystalline, vishwanath:crystalline}. Consequently, the landscape of TIs fragments according to the underlying band inversion momentum (${\bf K}_{\rm inv}$). However, boundary modes cannot distinguish them as they always exist at the interfaces of topological crystals, irrespective of ${\bf K}_{\rm inv}$.

\begin{figure}[b!]
\includegraphics[width=1.00\linewidth]{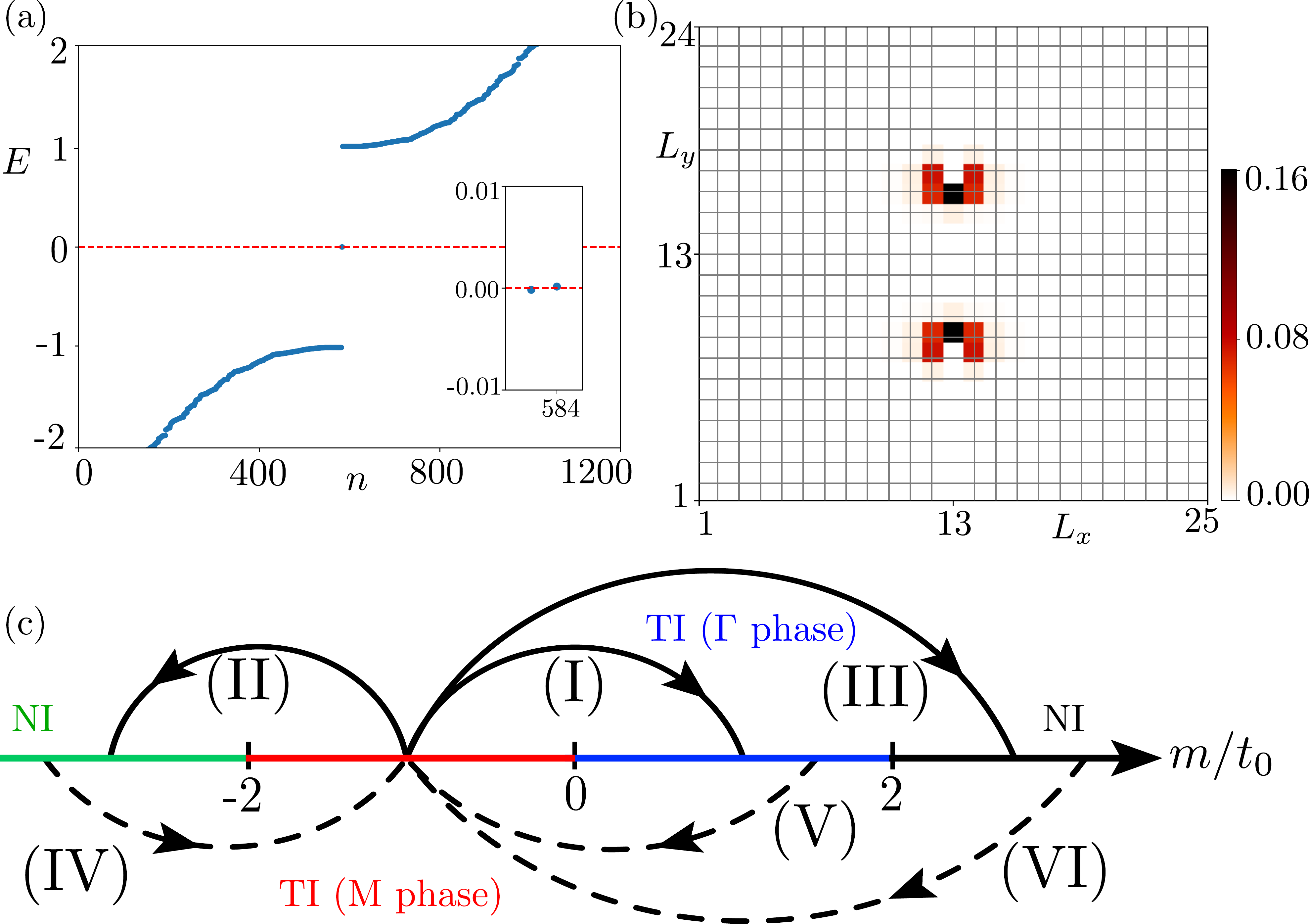}
\caption{(a) Energy spectra of the static Hamiltonian [Eq.~(\ref{eq:hamiltonian})] for $t=t_0=-m_0=1$, yielding a TATI with the band inversion at the ${\rm M}$ point (${\rm M}$ phase), in the presence of an edge dislocation-antidislocation pair with the periodic boundary condition in the $x$ and $y$ directions. The system then supports a pair of zero energy modes (inset). (b) The local density of states (LDOS) of these two modes are highly localized near the defect cores. (c) Phase diagram of the static Hamiltonian. Ramps out of (into) the ${\rm M}$ phase to (from) translationally inert insulators are shown by solid (dashed) arrows labeled by Roman numerals. The corresponding melting and condensation of dislocation modes are shown in Figs.~\ref{fig:fromMphasePureMixed}-\ref{fig:dislocationrecover}. Here, TI (NI) corresponds
to topological (normal) insulator.
}~\label{fig:Setup}
\end{figure}

Bulk topological lattice defects, such as dislocations, being sensitive to ${\bf K}_{\rm inv}$, are instrumental in distinguishing TIs. As dislocations are characterized by the nontrivial Burgers vector (${\bf b}$), electrons encircling the defect core picks up a hopping phase $\Phi_{\rm dis}={\bf K}_{\rm inv} \cdot {\bf b}$~\cite{ran-zhang-vishwanath, teo-kane, juricic:PRL, nagaosa, hughes-yao-qi, juricic:PRB, you-cho-hughes, cosma-quiroz-cano, roy-juricic-dislocation, nag-roy:floquetdislocation, panigrahi-moessner-roy:NHdislocation, velury-hughes, sanjibcosma:dislocation, panigrahi:commphy, salib-juricic-roy}. Evidently $\Phi_{\rm dis}=0$ in the $\Gamma$ phase, as ${\bf K}_{\rm inv}=0$ therein. If, on the other hand, ${\bf b}$ and ${\bf K}_{\rm inv}$ are such that $\Phi_{\rm dis}=\pi$ (modulo $2 \pi$), a nontrivial $\pi$ hopping phase threading the defect core binds localized gapless topological electronic modes therein (Fig.~\ref{fig:Setup})~\cite{juricic:PRL, juricic:PRB}, which have also been observed in experiments~\cite{disexp:1, disexp:2}. As dislocations are associated with the breaking of the local translational symmetry in the bulk of crystals, TIs harboring such defect modes are named translationally active topological insulators (TATIs). This general principle is applicable to two- and three-dimensional static and Floquet TIs and superconductors~\cite{ran-zhang-vishwanath, teo-kane, juricic:PRL, nagaosa, hughes-yao-qi, juricic:PRB, you-cho-hughes, cosma-quiroz-cano, roy-juricic-dislocation, nag-roy:floquetdislocation, panigrahi-moessner-roy:NHdislocation, velury-hughes, sanjibcosma:dislocation, panigrahi:commphy, salib-juricic-roy, disexp:1, disexp:2}.

\section{Broad questions and Key results}

Although unexplored thus far, with the recent progress at the frontier of dynamic topological phases, such as the ones realized in periodically driven Floquet materials~\cite{Flq:Thr1, Flq:Thr2, Flq:Thr3, Flq:Thr4, Flq:Thr5, Flq:exp1, Flq:exp2, Flq:exp3, Flq:exp4, Flq:exp5, Flq:exp6, Flq:exp7, Flq:exp8, Flq:exp9}, for example, the role of topological lattice defects in the dynamic realm arises as a timely issue of fundamental importance. In this context, we provide affirmative answers to the following questions. (a) Does the signature of topological dislocation modes survive in translationally inert insulators, reached from a TATI via a real time ramp? (b) Even more intriguingly, can topological dislocation modes be dynamically generated via a ramp, taking the system into a TATI phase from translationally inert insulators? 

\begin{figure}[t!]
\includegraphics[width=1.00\linewidth]{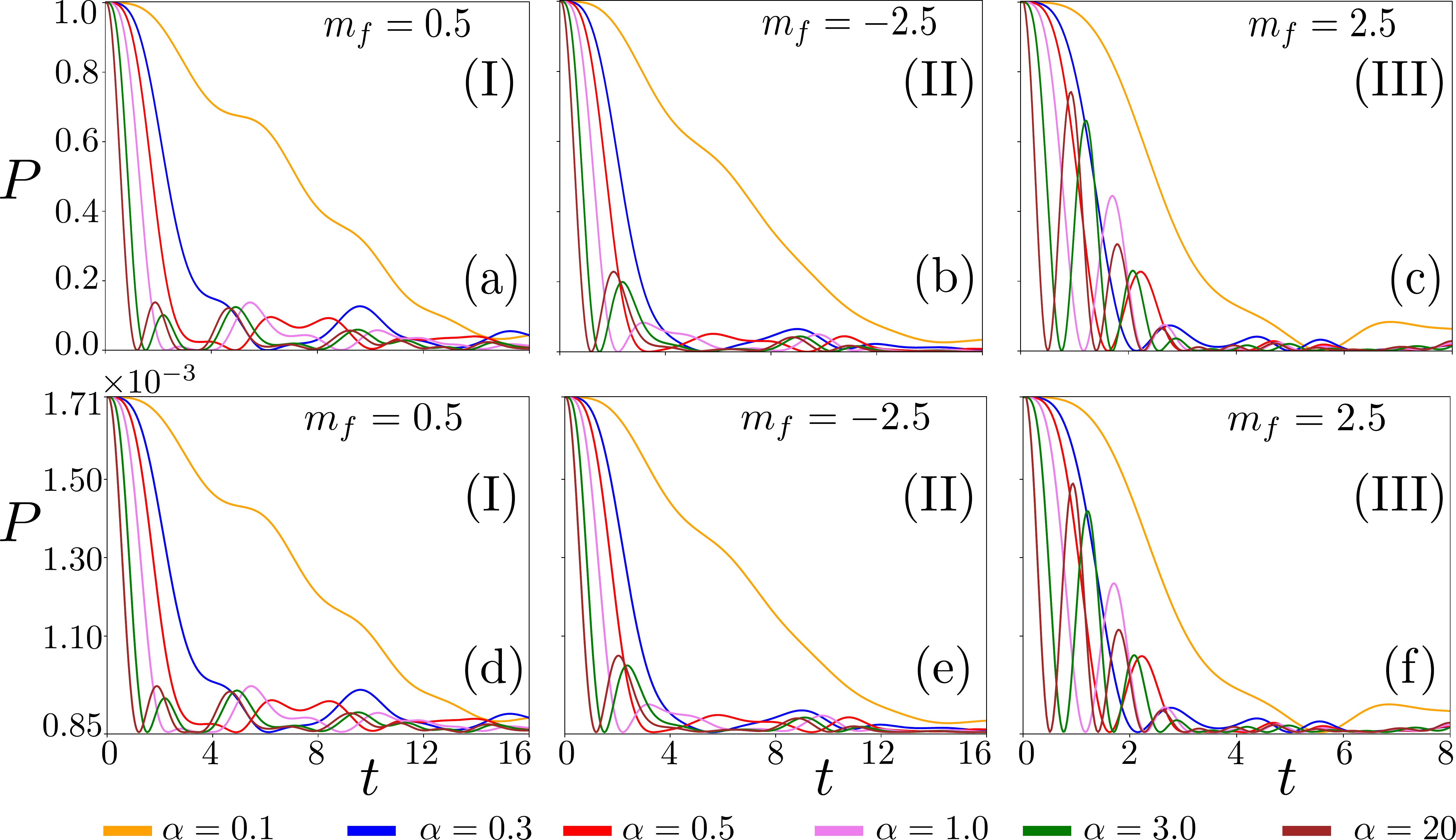}
\caption{Time evolution of the probability $P (t)$ [Eq.~(\ref{eq:probability})] of finding the dislocation mode in the presence of a real time ramp [Eq.~(\ref{eq:ramprofile})] that takes the system from the ${\rm M}$ phase (with $m_i=-1$) to a translationally inert TI with band inversion at the $\Gamma$ point [(a) and (d)] or normal insulator close to the ${\rm M}$ phase [(b) and (e)] or $\Gamma$ phase [(c) and (f)]. At $t=0$ the state is \emph{pure} (top), composed of a single dislocation mode or \emph{mixed} (bottom) with $N+1$ occupied single particle states (two dislocation modes and $N-1$ bulk states) of $H$ [Eq.~(\ref{eq:hamiltonian})], named the HF$^\prime$ state, for various choices of the ramp speed $\alpha$. Here $N$ is the total number of sites in a square lattice system in the presence of a dislocation-antidislocation pair. Therefore, signatures of the dislocation modes survive for a long time. The Roman numeral in each panel corresponds to the arrow out of the ${\rm M}$ phase shown in Fig.~\ref{fig:Setup}.  
}~\label{fig:fromMphasePureMixed}
\end{figure}

\begin{figure}[t!]
\includegraphics[width=1.00\linewidth]{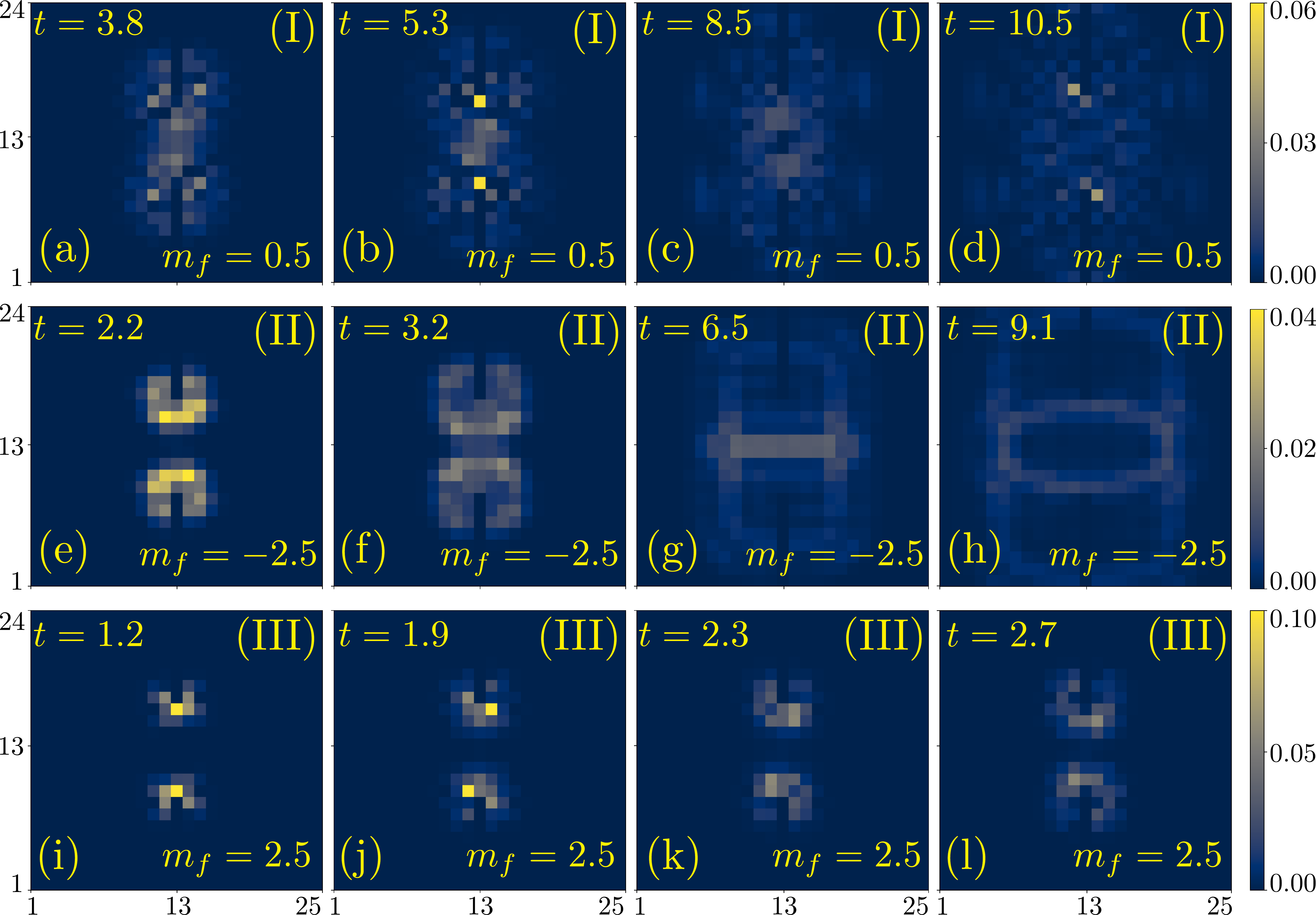}
\caption{Time evolution of the site resolved LDOS [Eq.~(\ref{eq:LDOS})], computed from the density matrix $\rho(t)$ at various time instants for a fixed ramp speed $\alpha=1$ [Eq.~(\ref{eq:ramprofile})]. Here, $\rho(0)$ is constructed from a single isolated dislocation mode for $t=t_0=-m_i=1$ (pure state). The final phase is a TI with the band inversion at the $\Gamma$ point [(a)-(d)], a normal insulator residing close to the ${\rm M}$ phase [(e)-(h)] or the $\Gamma$ phase [(i)-(l)]. The value of $m_f$ is quoted in each panel. Each row demonstrates dynamic melting of the localized dislocation mode due to the ramp. Compare with Fig.~\ref{fig:Setup}(b) and the color scale therein. Such time evolutions lead to valleys (peaks) in the probability of finding the initial dislocation modes when the LDOS appears prominently away from (near) the core of the lattice defect, as shown in the first and third (second and fourth) columns of each row. See Figs.~\ref{fig:fromMphasePureMixed}(a)-~\ref{fig:fromMphasePureMixed}(c) for the complete time evolution of these modes. We implement the lattice geometry shown in Fig.~\ref{fig:Setup}(b). Results are identical in the mixed HF$^\prime$ state once the uniform background LDOS for the half-filled system is subtracted. The Roman numerals in each panel corresponds to the arrow out of the ${\rm M}$ phase shown in Fig.~\ref{fig:Setup}.            
}~\label{fig:LDOSEvolution}
\end{figure}

\begin{figure*}[t!]
\includegraphics[width=1.00\linewidth]{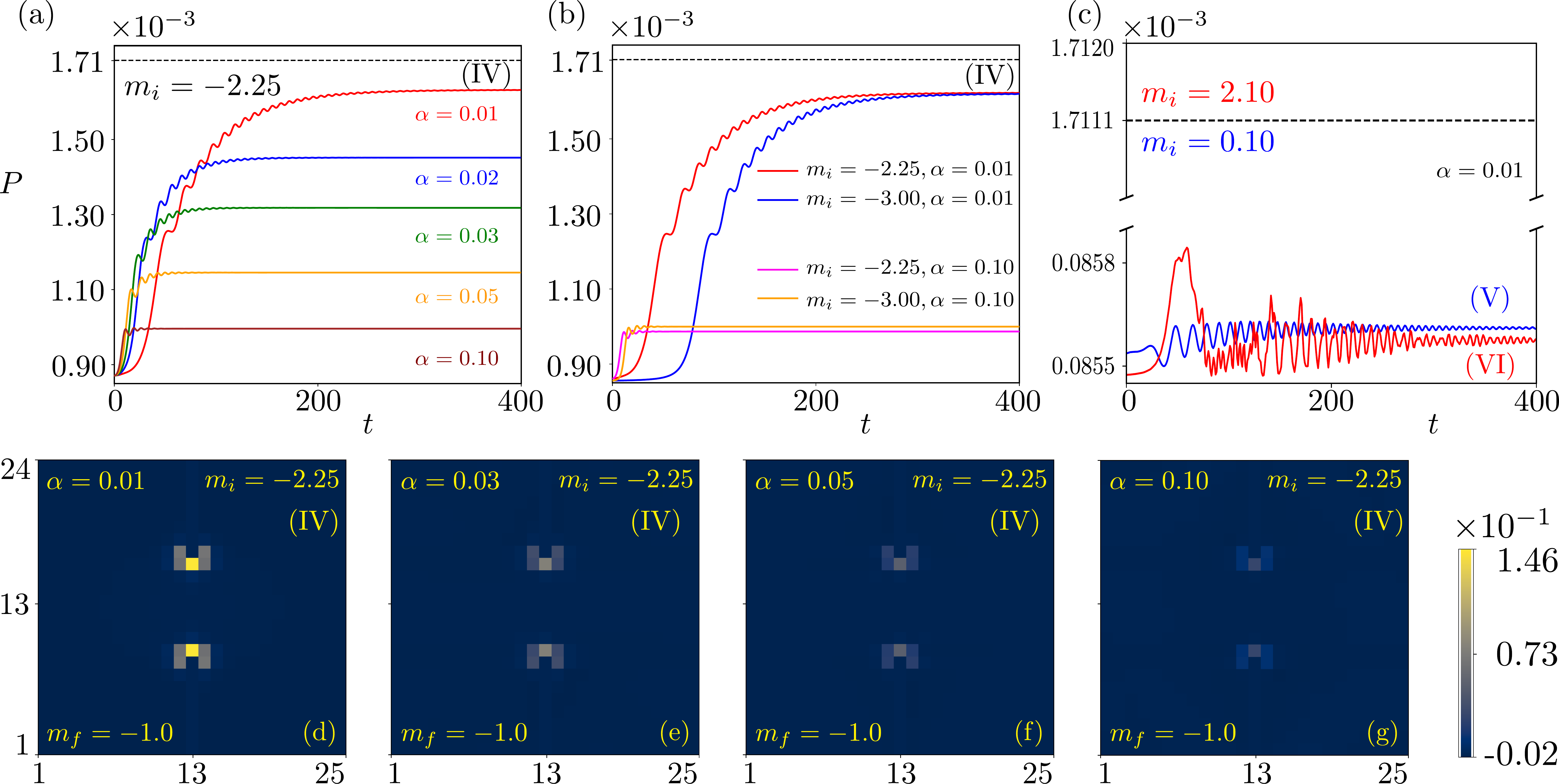}
\caption{Top: Probability $P(t)$ of dynamic condensation of dislocation modes. (a) Variation of $P(t)$ when the initial phase is a normal insulator residing close to the ${\rm M}$ phase for various ramp speeds $\alpha$. (b) Variation of $P(t)$ for two different band gaps (set by $m_i$) of the initial phase [same as in (a)] for two specific choices of $\alpha$. (c) Same as (a) but for a specific $\alpha$ and when the initial state features band minima (inverted for $m_i=0.10$ or noninverted for $m_i=2.10$) near the $\Gamma$ point. The dahed line corresponds to the maximal probability of finding dislocation modes in the mixed state, denoted by HF$^\prime$. Bottom: (d)-(g) Difference between the initial and final site resolved LDOS for various $\alpha$, confirming prominent dynamic generation of the dislocation modes, comparable to those in the ${\rm M}$ phase in the static system [Fig.~\ref{fig:Setup}(b)], for sufficiently slow ramp. The Roman number in each panel corresponds to the arrow into the ${\rm M}$ phase shown in Fig.~\ref{fig:Setup}.       
}~\label{fig:dislocationrecover}
\end{figure*}

To answer these questions, we subscribe to a paradigmatic lattice model for two-dimensional time-reversal symmetry breaking insulators. Besides featuring the translationally active ${\rm M}$ phase with the band inversion at the ${\rm M}=(\pi,\pi)/a$ point of the BZ, it also accommodates translationally inert normal or trivial insulators (with no band inversion) as well as a TI with the band inversion at the $\Gamma=(0,0)$ point [Fig.~\ref{fig:Setup}(c)]. Here $a$ is the lattice constant of an underlying square lattice. Then from the time ($t$) evolution of the appropriate density matrix $\rho(t)$, governed by the von Neumann equation
\allowdisplaybreaks[4] 
\begin{equation}~\label{eq:heisenberg}
\frac{d \rho(t)}{dt}=-\frac{i}{\hbar} \; \left[ H(t), \rho(t) \right],
\end{equation} 
where the Hamiltonian $H(t)$ captures the real time ramp, we make the following key observations. When the system is initially prepared in a TATI phase with a pair of dislocation modes and the ramp takes it into one of the translationally inert phases, gradually decaying signatures of the defect modes survive for a long time. This outcome is qualitatively insensitive to the nature of the final state. See Figs.~\ref{fig:fromMphasePureMixed} and \ref{fig:LDOSEvolution}. By contrast, when the ramp takes a reverse course, taking the system into the ${\rm M}$ phase from one of the translationally inert insulators, topological dislocation modes dynamically condense near the defect cores. The probability of such dynamic generation of the topological defect modes increases with decreasing ramp speed, resembling the \emph{adiabatic theorem}. However, dynamic nucleation of the defect modes is most prominent when the ramp begins from a normal insulator, residing close to the TATI or ${\rm M}$ phase. These findings are showcased in Fig.~\ref{fig:dislocationrecover}. Throughout this paper we set $\hbar=1$.

\section{Model}

The Hamiltonian for the lattice model is~\cite{qiwuzhang} 
\allowdisplaybreaks[4]
\begin{equation}~\label{eq:hamiltonian}
H=t_1 \sum_{j=x,y} \sin(k_j a) \tau_j + \bigg( t_0 \sum_{j=x,y} \cos(k_j a) -m_0 \bigg) \tau_z.
\end{equation}       
The vector Pauli matrix ${\boldsymbol \tau}=(\tau_x,\tau_y,\tau_z)$ operates on the orbital indices. Translationally active ${\rm M}$ and inert $\Gamma$ topological phases are realized for $-2<m/t_0<0$ and $0<m/t_0<2$, respectively. Otherwise the system is a normal insulator. See Supplemental Material~\cite{supplementary}. Only the ${\rm M}$ phase supports a pair of near zero energy dislocation modes that are highly localized around the defect cores, when ${\bf b}=a {\bf e}_x$, for example (Fig.~\ref{fig:Setup}). We denote them as $\ket{\Psi^{\rm dis}_i}$ with $i=1,2$. They are related via an antiunitary particle-hole symmetry $\ket{\Psi^{\rm dis}_{1,2}}=\Theta \ket{\Psi^{\rm dis}_{2,1}}$ up to an overall unimportant phase, where $\Theta=\tau_x {\mathcal K}$ and ${\mathcal K}$ is the complex conjugation, as $\{H, \Theta \}=0$~\cite{royantiunitary}.

The real time ramp under which $m_0 \to m(t)$, where  
\allowdisplaybreaks[4]
\begin{equation}~\label{eq:ramprofile}
m(t)= m_i+ \left( m_f-m_i \right) \: \left[ 1- \exp(-\alpha t) \right]
\end{equation} 
is characterized by the ramp speed $\alpha$. This ramp profile with $m(0)=m_i$ and $m(t \to \infty)=m_f$, where the subscript $i$ ($f$) stands for initial (final), allows the system to interpolate between any two insulators appearing in the phase diagram of $H$ through band gap closing with tunable ramp speed~\cite{budich-zoller}. This ramp does not break any lattice symmetry. When $m_0 \to m(t)$ in Eq.~(\ref{eq:hamiltonian}), we denote the time dependent Hamiltonian by $H(t)$ [Eq.~(\ref{eq:heisenberg})].

\section{Results}

We set the stage with the discussion of a simple situation. Consider an isolated \emph{pure state} $\ket{\Psi^{\rm dis}_1}$, for which the density matrix $\rho(0)= \ket{\Psi^{\rm dis}_1} \bra{\Psi^{\rm dis}_1}$. Under the time evolution this state always remains pure. Once the ramp [Eq.~(\ref{eq:ramprofile})] is switched on, at any given instant of time the probability of finding the dislocation mode is 
\allowdisplaybreaks[4]
\begin{equation}~\label{eq:probability}
P(t)=\bra{\Psi} \rho(t) \ket{\Psi},
\end{equation} 
also known as the \emph{fidelity}, but only when $\rho(0)$ and thus $\rho(t)$ represents a pure state, where $\ket{\Psi} = \ket{\Psi^{\rm dis}_1}$, and $\rho(t)$ is obtained by numerically solving Eq.~(\ref{eq:heisenberg}). The results are shown in Figs.~\ref{fig:fromMphasePureMixed}(a)-~\ref{fig:fromMphasePureMixed}(c). Irrespective of the nature of the final translationally inert insulator (topological or normal), we find that $P(t)$ remains appreciably finite for a long time over a wide range of $\alpha$. For a small $\alpha$, $P(t)$ decays slowly, as the system then takes a longer time to escape the ${\rm M}$ phase. For faster ramp $P(t)$ initially falls rapidly. However, it then shows oscillatory behavior for a long time. These characteristics of $P(t)$ for sufficiently large $\alpha$ are (almost) identical to those for the survival probability of a single dislocation mode across a sudden quench~\cite{supplementary, survivalbook, nag-juricic-roy}. The time evolution of $P(t)$ reflects on the time dependence of the site resolved LDOS, computed from the density matrix as
\begin{equation}~\label{eq:LDOS}
D_i(t)=\sum_{\tau=1,2} \bra{i,\tau} \rho(t) \ket{i,\tau},
\end{equation} 
where $i$ ($\tau$) is the site (orbital) index, and $\ket{i,\tau}$ is the single particle state vector at site $i$ with orbital $\tau$. The results are shown in Fig.~\ref{fig:LDOSEvolution}. Irrespective of the nature of the final insulator, we find that the LDOS spreads over the entire system with increasing time causing the overall decay of $P(t)$ with $t$. Nonetheless, the LDOS displays periodic peaks and deeps at the dislocation core, where the initial defect mode $\ket{\Psi^{\rm dis}_1}$ was prominently localized, resulting in an oscillatory behavior of $P(t)$, shown in Figs.~\ref{fig:fromMphasePureMixed}(a)-\ref{fig:fromMphasePureMixed}(c).

However, in quantum materials, midgap dislocation modes can only be occupied upon filling all the negative-energy bulk states. Therefore, to experimentally observe the dynamic melting of defect modes, one needs to consider an appropriate many-body ground state. Due to the particle-hole symmetry, a half-filled system displays a uniform average electronic density equal to \emph{one} in the entire system irrespective of the presence or absence of dislocation modes at any time for any $\alpha$~\cite{supplementary}. To capture the time evolution of the dislocation modes, we therefore need to add one fermion to the half-filled sea, a state denoted by HF$^\prime$. The density matrix for this \emph{mixed} state in a system containing $N$ number of sites reads 
\allowdisplaybreaks[4]
\begin{equation}~\label{eq:densitymatrixmixed}
\rho(0)= \frac{1}{N+1} \sum^{N+1}_{i=1} \ket{\Psi_i} \bra{\Psi_i},
\end{equation}     
where $\ket{\Psi_i}$ is an eigenstate of $H(0)$ with energy $E_i$. Evidently $\rho(0)$ cannot be expressed as $\ket{\Psi} \bra{\Psi}$ in terms of some state vector $\ket{\Psi}$ as it is a mixed state~\cite{ballentine}. In HF$^\prime$, two dislocation modes are occupied, besides $N-1$ number of bulk states. Then the LDOS shows a sharp peak at the dislocation core in the ${\rm M}$ phase once the average background LDOS for the half-filled system is subtracted.

The time evolution of this density matrix is obtained by numerically solving Eq.~(\ref{eq:heisenberg}), from which we compute the probability of finding the dislocation mode at $t \geq 0$ from Eq.~(\ref{eq:probability}) with $\ket{\Psi}=(\ket{\Psi^{\rm dis}_1}+\ket{\Psi^{\rm dis}_2})/\sqrt{2}$. The results are shown in Figs.~\ref{fig:fromMphasePureMixed}(d)-~\ref{fig:fromMphasePureMixed}(f), which are \emph{identical} to the ones we obtained for the pure state [Figs.~\ref{fig:fromMphasePureMixed}(a)-~\ref{fig:fromMphasePureMixed}(c)], except that $P(0)=(N+1)^{-1}$ since all the $(N+1)$ filled states in HF$^\prime$ are equally probable to be occupied initially. And min.$\{ P(t)\}=1/(2N)$, which corresponds to a maximally disordered state (all $2N$ single-particle states are equally probable to be occupied) with maximal von Neumann entropy~\cite{supplementary}. The site resolved LDOS shows identical behavior to that in Fig.~\ref{fig:LDOSEvolution} for the pure state once the background of the uniform LDOS for the half-filled state is subtracted.

Finally, we showcase the dynamic buildup of dislocation modes when the real time ramp brings the system, initially prepared in one of the translationally inert insulating phases, to the translationally active ${\rm M}$ phase. Following the discussion from the last paragraph, we immediately recognize that the initial system must be prepared in the HF$^\prime$ state, which is now devoid of any defect mode, with the density matrix $\rho(0)$ [Eq.~(\ref{eq:densitymatrixmixed})]. From $\rho(t)$, we then compute the probability of finding the dislocation modes [Eq.~(\ref{eq:probability})] and site resolved LDOS [Eq.~(\ref{eq:LDOS})] at any instant of time. Results are shown in Fig.~\ref{fig:dislocationrecover}. Notice that the initial system is devoid of any defect modes and all the filled states therein contribute in the dynamic condensation of dislocation modes via the real time evolution. Hence, no pure state can capture this phenomenon and we always have to work with the mixed states.

Dynamic condensation of dislocation modes is most prominent when the initial state is a normal insulator, residing close to the ${\rm M}$ phase, with the noninverted band minima near the ${\rm M}$ point. Recall that dislocation modes originate from the band inversion at the ${\rm M}$ point (${\bf K} \cdot {\bf b}$ rule), which is easier to achieve dynamically when the minima of noninverted bands is near the ${\rm M}$ point initially, as this process requires a small momentum transfer. Otherwise, with slower ramp speed the probability of dynamic generation of the dislocation modes increases [Fig.~\ref{fig:dislocationrecover}(a)], which is insensitive to the magnitude of the initial band gap [Fig.~\ref{fig:dislocationrecover}(b)]. These observations can be reconciled with the adiabatic theorem: For adiabatically slow ramp, the system should always find itself in the instantaneous ground state. As a consequence, for a sufficiently slow ramp, the hallmark dislocation modes of the translationally active ${\rm M}$ phase develop more prominently as $t \to \infty$. These conclusions are further substantiated from the difference between initial and final LDOS at each site for various $\alpha$, shown in Fig.~\ref{fig:dislocationrecover}(d)-(g). Remarkably, the LDOS near the dislocation cores recovers more than 95\% of the weight of the original dislocation modes in the ${\rm M}$ phase of the static system for a sufficiently slow ramp ($\alpha=0.01$). Such recovery of dislocation modes, however, gets weaker with increasing ramp speed. By contrast, when the system is initially prepared in an insulating phase, featuring minima of inverted or noninverted bands near the $\Gamma$ point, the probability of dynamic generation of the dislocation modes remains negligibly small even for sufficiently slow ramp [Fig.~\ref{fig:dislocationrecover}(c)], as the requisite dynamic band inversion at the ${\rm M}$ point now demands a \emph{large} momentum transfer.

All the results presented here are expected to mimic the outcomes in the thermodynamic limit, as for sufficiently large number of lattice sites ($N$) in the system, with $N=1176$ in our setup, the exact numerical diagonalization produces (a) the same band gap predicted from the Bloch Hamiltonian [Eq.~\eqref{eq:hamiltonian}] and (b) topological defect modes in the ${\rm M}$ phase at energy equal to zero within the numerical accuracy. See Fig.~\ref{fig:Setup}(a).

\section{Discussions and outlooks}

Defects are ubiquitous in crystals. Here we outline a theoretical framework to capture dynamic melting and condensation of topological dislocation modes when a real time ramp takes the system out of and into a TATI, respectively. The proposed formalism based on the time dynamic of an appropriate density matrix governed by the von Neumann equation is sufficiently general that it can be extended to any topological system, including topological superconductors, in an arbitrary dimension and belonging to arbitrary symmetry class to capture similar dynamic phenomena for boundary (such as edge, surface, hinge, and corner) as well as defect (line dislocation and grain boundaries, for example) modes. Our model Hamiltonian also describes a $p_x+ip_y$ superconductor with amplitude $t_1$, and the band inversion near the $\Gamma$ and ${\rm M}$ points manifests Fermi surfaces near them. Then a weak coupling topological BCS pairing sets in and dislocation cores host localized Majorana modes. Tunable time dynamics of localized Majorana modes can in principle be useful in quantum computation and information technologies, facilitating information storage over a desired time. These fascinating avenues will be explored in future. Meanwhile, it will be worthwhile to investigate the impact of \emph{dephasing} in these open quantum systems, which may boost the revival probability of dislocation modes when its dynamic generation requires a large momentum transfer. Results reported in this paper suggest a dynamic approach to distinguish two normal insulators, residing close the TIs with band inversion at the $\Gamma$ and ${\rm M}$ points, from the revival probability of the defect modes when time-ramped into the ${\rm M}$ phase, thus complementing a recent proposal to distinguish such normal or atomic insulators by inducing superconductivity therein~\cite{dasmannaroy2023}.

The requisite time-modulated band gap in quantum crystals and cold atomic lattices can, for example, be accomplished by time-dependent hydrostatic pressure and on-site staggered-orbital potential, respectively, allowing a smooth interpolation between topologically distinct insulators. The predicted dynamic melting and more intriguing condensation of dislocation modes can be detected from the time-dependent LDOS, measured near the defect cores by scanning tunneling spectroscopy. In metamaterials, the time-dependent band gap can be engineered by (a) coupling electric nodes containing capacitors and/or inductors with resistors in active topolectric circuits to minimize gain and loss, such that the circuit decay constant provides the desired profile of $m(t)$~\cite{thomale:time} or (b) a time-dependent ratio of the distance between intra unit-cell magnetomechanical resonators (optical waveguides) to the effective lattice constant of mechanical (photonic) lattices. In these metacrystals, such ratios have been tuned to trigger transitions between topological and normal insulators, and lattice defects as well have been engineered, featuring robust defect modes~\cite{photonic:1, photonic:2, photonic:3, photonic:4, mechanical:1}. The proposed dynamic responses of defects can be measured from the time-dependent impedance (in topolectric circuits), mechanical susceptibility (in mechanical systems), and two-point pump probe (in photonic lattices), all of which mimic the LDOS. With the recent progress in realizing Floquet topological phases in quantum materials~\cite{Flq:exp1, Flq:exp2, Flq:exp3} and metamaterials~\cite{Flq:exp4, Flq:exp5, Flq:exp6, Flq:exp7, Flq:exp8, Flq:exp9}, the ramp dynamics and dynamic buildup of topological defect modes should be within the reach of current experimental facilities.

\acknowledgments

S.K.D.\ was supported by a Startup grant of B.R. from Lehigh University. B.R.\ was supported by NSF CAREER Grant No.\ DMR- 2238679. We thank Suvayu Ali for technical support.

\end{document}